\documentclass[twocolumn,pre,aps,showpacs,amsmath,amssymb]{revtex4-1}
\usepackage{graphicx}
\usepackage{bm}
\usepackage{epstopdf}
\begin{document}

\title{Stochastic thermodynamics of non-harmonic oscillators in high vacuum}
\author{Domingos S. P. Salazar$^{1}$}
\email[]{salazar.domingos@gmail.com}
\author{S\'ergio A. Lira$^{2}$}
\email[]{sergio@fis.ufal.br}
\affiliation{$^1$ Unidade Acad\^{e}mica de Educac\~{a}o a Dist\^{a}ncia e Tecnologia, Universidade Federal Rural de Pernambuco, Recife, Pernambuco 52171-900 Brazil \\
$^2$ Instituto de F\'{\i}sica, Universidade Federal de Alagoas, Macei\'o, Alagoas 57072-900 Brazil}


\begin{abstract}
We perform an analytic study on the stochastic thermodynamics of a small classical particle trapped in a time dependent single-well potential in the highly underdamped limit. It is shown that the nonequilibrium probability density function for the system's energy is a  Maxwell-Boltzmann distribution (as in equilibrium) with a closed form time dependent effective temperature and fractional degrees of freedom. We also find that the solvable model satisfies the Crooks fluctuation theorem, as it is expected. Moreover, we compute the average work in this isothermal process and characterize analytically the optimal protocol for minimum work. The optimal protocol presents an initial and a final jumps which correspond to adiabatic processes linked by a smooth exponential time dependent part for all kinds of single-well potentials. Furthermore, we argue that this result connects two distinct relevant experimental setups for trapped nano-particles: the levitated particle in a harmonic trap, and the free particle in a box; as they are limiting cases of the general single-well potential and display the time-dependent optimal protocols. Finally, we highlight the connection between our system and an equivalent model of a gas of Brownian particles.
\end{abstract}
\maketitle

\section{Introduction}

Modeling nonequilibrium physics of very small systems requires the definition of thermodynamic quantities such as heat and work for single particle trajectories \cite{Seifert2012,Bustamante2005,Sekimoto2010}. Since the original Brownian motion, stochastic thermodynamics has further evolved with the development of optical tweezers, which allowed precise trapping and cooling procedures for levitated nanoparticles and started the field of optomechanics \cite{Aspelmeyer2014,Gieseler2018}.
Applications include the experimental verification of fluctuation theorems (FTs) in biophysics at molecular level \cite{Liphardt2002,Collin2005,Alemany2012}, apparent violation of the second law of thermodynamics \cite{Wang2002} and evidence of Landauer's principle \cite{Gavrilov2014}. More recently, in contrast to suspended particles, experimental groups obtained measurements of 
particles optically trapped in high vacuum and verified several nonequilibrium results, including the ubiquitous FTs \cite{Jarzinski1997,Yin2013,Vamivakas2016,Gieseler2012,GieselerPRL2012,Hoang2018,Millen2015,Tongcang2011}. This experimental frontier might provide an interface to test quantum limits \cite{Jain2016}, non-Newtonian \cite{Geraci2010} and quantum gravity \cite{Bose2017}, and the realization of feasible underdamped nanomechanical heat engines \cite{Dechant2015}.

In this context, research is devoted to design and implement efficient thermal engines at micro- and nanoscale \cite{Blickle2011, Abah2012, Rossnagel2014,Dechant2017}, which includes optimal protocols for producing minimum average work over finite time windows. In optical traps, a protocol might be obtained adjusting the laser trap frequency, which is equivalent to change the stiffness of the restoring force. How the frequency is tuned over a finite time window produces a variety of nonequilibrium thermodynamic processes, resulting in fluctuating work and heat. In this sense, optimal protocols for a time-dependent driving force in overdamped systems have been extensively studied and found to exhibit discontinuous jumps in some situations \cite{Schmiedl2007,Schmiedl2008,Aurell2011,Aurell2012}, where the trap frequency is required to change abruptly, followed by a smooth tuning. The same interesting feature was observed in underdamped models for optical traps in numerical simulation \cite{Gomez2008,Dechant2015} and also in analytical calculations \cite{Dechant2017}. Besides the adiabatic jumps, the continuous part of the optimal protocols in such underdamped systems is exponential in time \cite{Dechant2017,Agarwal2013}. Surprisingly, the same exponential protocol was also obtained for a free particle in a box \cite{GONG16} using different methods. How general is this exponential behavior and the presence of jumps in optimal protocols is a question that remains open in stochastic thermodynamics.

In this work, we propose a solvable model for a levitated particle in the highly underdamped limit for a general single-well potential of the type $U(x)\propto x^{2n}$. Our result, which is of special experimental interest \cite{Gieseler2018,Vamivakas2016}, generalizes previous results limited to the harmonic case ($n=1$) \cite{Dechant2017} and the particle in a box ($n\rightarrow \infty$) \cite{GONG16}, as well as consider finite time processes in the presence of damping, expanding previous analysis of single-well potentials \cite{Mallick2003, Mallick2005, Mandrysz2018}. We show the model has a closed form propagator for the stochastic energy and simple expressions for averaged thermodynamic quantities. Remarkably, we find the nonequilibrium probability distribution for the energy in a isothermal process is a Maxwell-Boltzmann (MB) distribution with a time dependent effective temperature. As an application, we study the optimal protocol that produce minimum work in a isothermal process and find finite jumps in the driving protocol, whose magnitude depends on $n$, combined with a time-dependent exponential relaxation for all $n$.

The analysis is organized as follows. We present a stochastic model for the mechanical energy in section II. Then, we show find the propagator for the energy in section III and show the model satisfies Crooks fluctuation theorem in section IV. As an application of the model, in section V, we find the optimal protocol that minimizes the average irreversible work over a finite time interval. In section VI, we present a analogy between the model and a gas of free Brownian particles. Finally, in Section VII, we present some conclusions and perspectives.

\section{The Stochastic Model} 
We consider the dynamics of a small Brownian  oscillator submitted to the external potential $U(x,k_t)=k_t x^{2n}/2n$, where $k_t$ is a time-dependent generalized stiffness which can be varied by tuning the frequency of the trapping setup. Here we introduce parameters $k_t=m\Omega_t^2 L^{2-2n}$, where $\Omega_t$ has units of frequency ($s^{-1}$) and $L$ is a characteristic length. The particle is in contact with a thermal reservoir of temperature $T$ and it is modelled using the Langevin equation
\begin{equation}
    \label{Langevin}
    \ddot{x} + \Gamma_0\dot{x} +\Omega_t^2 L \big(\frac{x}{L}\big)^{2n-1} = \frac{1}{m}F_{fluc}(t),
\end{equation}
for the particle position $x(t)$. The random Langevin force $F_{fluc}(t)$ is normally distributed with zero mean and its components satisfy $\langle F_{fluc}(t)F_{fluc}(t')\rangle = 2m\Gamma_0 k_B T \delta(t-t')$, where $\Gamma_0$ is a friction coefficient \cite{Gieseler2012} and $m$ is the particle mass. Considering the usual scale of observation and the suspension media involved, the system is commonly solved in the overdamped limit \cite{Sekimoto2010}, where the inertia term is neglected, $m\ddot{\textbf{x}}(t)\approx0$. For levitated particles in highly diluted media \cite{Gieseler2012,Gieseler2018}, although the mass of the particles is small, this regimen imposes a small friction coefficient $\Gamma_0$ compared to the frequency $\Omega_t$, so inertial effects cannot be neglected. Here we show that using this last condition in (\ref{Langevin}) results in a solvable stochastic differential equation (SDE) for the system's total energy given by $E(x,p)=p^2/2m+k_t x^{2n}/2n$, with momentum $p=m\dot{x}$. After a suitable combination of Ito's Lemma \cite{Oeksendal2003} and the highly underdamped limit ($\Omega_t\gg\Gamma_0$) \cite{Gieseler2012,Paper01,Mallick2003}, the energy SDE is given by
\begin{equation}
\label{LangevinforE1}
dE=d'W -\Gamma_n (E-\frac{f_n}{2}k_B T)dt+\sqrt{2\Gamma_n k_BTE}d\textrm{B}_t,
\end{equation}
where 
\begin{equation}
\label{infW}
d'W=\frac{\partial U(x,k)}{\partial k} \dot{k_t}dt
=\frac{\dot{k_t}}{k_t}U(x,k_t)dt
\end{equation}
is the increment of work \cite{Sekimoto2010}, $d\textrm{B}_t$ is the increment of the Wiener process with a redefined friction coefficient
\begin{equation}
    \label{gamman}
    \Gamma_n=\frac{2n}{n+1}\Gamma_0,
\end{equation}
ranging from $\Gamma_0$, in the case of harmonic potential, to $2\Gamma_0$, in the case of a particle in a box. The system also presents effective degrees of freedom $f_n$ given by
\begin{equation}
    \label{fn}
    f_n=\frac{n+1}{n},
\end{equation}
ranging from $f_1=2$, consistent with the harmonic potential in one dimension \cite{Gieseler2012}, to the minimum $f_\infty=1$, also consistent to a particle in a box in one dimension (ie, a single degree of freedom). Notice that intermediate potentials ($1<n<\infty$) have fractional $f_n$. In such potentials, the (average) kinetic energy is a $n$ dependent fraction of the total mechanic energy, given by the virial theorem, and this information is encoded in $f_n$. We remark that the thermal coupling in (\ref{Langevin}) could be generalized to a correlated stochastic force, where similar expressions for the work and heat functional can be obtained \cite{Pal2014}, within the validity of the highly underdamped limit approximation. In such cases, introducing a stochastic drive affects the statistics of work and heat, as well as their underlying optimal protocols. In this paper, we limit the scope of (\ref{Langevin}) to purely thermal noise, as it consists on a relevant experimental setup \cite{Gieseler2019}.

In order to make use of the SDE (\ref{LangevinforE1}) in the highly underdamped limit, we define three separated timescales explicitly: 
\begin{equation}
\label{timescales}
\delta_i\ll \tilde{\tau} \ll \Gamma_n^{-1}.
\end{equation}
The smallest timescale is the trap period $\delta_i=2\pi \Omega_i^{-1}$ at time $t_i$. All finite protocols happen over a timescale $\tau \geq \tilde{\tau}$, where $\tilde{\tau}$ is the smallest timescale for a protocol (called jump protocols). The largest timescale is the thermal $\Gamma_n^{-1}$, where dissipation effects are observed. Notice that jump protocols seem abrupt in the thermal timescale, as $\tilde{\tau}\Gamma_n \approx 0$. Alternatively, smooth protocols are observed over a large time interval, with $\tau \Gamma_n>0$.

Fast transformations such that $\tilde{\tau} \approx \delta_i$ are beyond the scope of this work. These requirements limit the applicability  of the current approach to systems with a very wide timescale separation between $\delta_i$ and $\Gamma_n^{-1}$. This is actually the case for some experiments. For instance, a recent experimental setup for levitated nanoparticles trapped in a laser uses $\delta_i^{-1} \approx 125$ kHz and $\Gamma_1^{-1} \approx 0.17 s$ \cite{Gieseler2012}. For such timescales separation, an interval $\tilde{\tau}\approx 10^{-3}s$ would result in $\delta_i /\tilde{\tau} < 10^{-2}$ and $\tilde{\tau}/\Gamma^{-1}< 10^{-2}$, which seems a suitable condition for (\ref{timescales}).

The effect of the timescales separation in the description of the energy (\ref{LangevinforE1}) is presented in the following subsections. First, we use the definition for $d'W$ from (\ref{infW}) to define the work over a small oscillation period, $dW$, for smooth protocols using the virial theorem. Finally, we derive the final SDE for the energy and aply the expression to compute the work in jump protocols, taking the limit $\tilde{\tau} \ll \Gamma_n^{-1}$.

\subsection{Work increment}
In this section, we use the large timescales separation (\ref{timescales}) in order to find an approximation for the work increment (\ref{infW}) averaged over a oscillation period. In this situation, the potential energy $U(x,k)$ is related to the mechanical energy by the virial theorem, $(1/\delta_i)\int U(x,k)dt=E/(n+1)$, as a generalization of the approximations previously used for the harmonic case ($n=1$) \cite{Gieseler2012,Dechant2017}. Therefore, we may integrate (\ref{infW}) over this small time interval, $\delta_i$, and find
\begin{equation}
\label{Workapprox}
\delta W=\int_t^{t+\delta_i}d'W\approx\frac{\dot{k_t}}{k_t}\int_t^{t+\delta t}U(x,k_t)dt=\frac{\dot{k_t}}{k_t}\frac{E}{n+1}\delta_i,
\end{equation}
which holds for slow protocols satisfying (\ref{timescales}), such that $k_{t+\delta_i} \approx k_t$. Taking the limit $\delta_i \rightarrow dt$, the approximation above leads to a closed form SDE for the energy discussed in the next section. We remark that the definition and treatment of adiabatic (jump) protocols is presented in Sec. II. C.

\subsection{SDE for the energy}
Finally, from the expression for the differential of work (\ref{Workapprox}), we obtain the increment of work over one oscillation, 
\begin{equation}
\label{diffW}
dW = \frac{\dot{k}(t)}{k(t)}\frac{E(t)}{n+1}dt.
\end{equation}
We remark that expression (\ref{diffW}) generalizes approximations previously used for the harmonic case ($n=1$) \cite{Gieseler2012,Paper01,Dechant2017}. Finally, we replace (\ref{diffW}) in (\ref{LangevinforE1}) and obtain the final SDE for the energy,
\begin{equation}
\label{LangevinforE2}
dE=-\Gamma_n ((1-\frac{\dot{\lambda_t}}{2\lambda_t\Gamma_n})E-\frac{f_n}{2}k_B T)dt+\sqrt{2\Gamma_n k_BTE}d\textrm{B}_t,
\end{equation}
where $\lambda_t=k_t^{2/(n+1)}$. We have omitted the index $n$ in $\lambda_t$ for clarity. The SDE for $E(t)$ becomes self contained with contributions from heat and work. It is worth noting that the SDE is identical for all potentials $n=1,2,...$ in terms of the adjusted parameters $\Gamma_n,f_n$ and $\lambda_t$, which makes the stochastic thermodynamics of all single-well potentials equivalent in the highly underdamped limit.

Several applications emerge from (\ref{LangevinforE2}) depending on the potential ($n$) and the protocol $\lambda_t$. Notice that, in the case $n=1$ (harmonic potential), the SDE (\ref{LangevinforE2}) reduces to the levitated particle trapped by a laser in the highly underdamped limit \cite{Dechant2017}. We will also argue that the case $n\rightarrow \infty$ models the particle in a box \cite{GONG16}. Notably, intermediate cases ($1<n<\infty$) are modeled by fractional degrees of freedom, $1<f_n<2$. In any case, for a constant protocol $\lambda_t=\lambda_0$ and $\dot{\lambda_t}=0$, the SDE describes the heat exchanged for a isochoric process ($dW=0$), where the heat distribution PDF has a closed form that resembles the PDF obtained in similarly to the case $n=1$ \cite{Paper01}. For completeness, we also show from (\ref{LangevinforE1}) that the approximate work (\ref{diffW}) satisfies the Crooks fluctuation theorem (CFT) 
\cite{Crooks1999,GONG15} for any $n$ in the next sections. As a first application we calculate the work of a adiabatic transformation.

\subsection{Application: Adiabatic protocols}
We define adiabatic processes (jumps) as the protocols over the time interval $\tilde{\tau}$ such that $\tilde{\tau} \ll \Gamma_n^{-1}$. In this limit, the energy SDE (\ref{LangevinforE2}) becomes a ordinary differential equation (ODE),
\begin{equation}
\label{ODE}
    dE = dW = \frac{E}{2}\frac{\dot{\lambda}}{\lambda}dt,
\end{equation}
since heat is negligible. The ODE (\ref{ODE}) has a simple solution,
\begin{equation}
\label{adiabaticenergy}
    E(t)=E_0(\frac{\lambda(t)}{\lambda_0})^{1/2}.
\end{equation}
Inserting the solution above in the definition of work (\ref{diffW}) results in
\begin{equation}
    W_{jump} = \frac{1}{2}\int_0^{\tilde{\tau}} E \frac{\dot{\lambda}}{\lambda}dt=\frac{E_0}{2\lambda_0^{1/2}}\int_0^{\tilde{\tau}}
    \lambda^{-1/2} \dot{\lambda_t} dt,
\end{equation}
and using $\lambda^{-1/2}\dot{\lambda}=2(d/dt)\lambda^{1/2}$, we obtain
\begin{equation}
    W_{jump} =\frac{E_0}{\lambda_0^{1/2}}\int_0^{\tilde{\tau}}
    \frac{d}{dt}\lambda^{1/2} dt=E_0(\sqrt{\frac{\lambda_1}{\lambda_0}}-1),
\end{equation}
which is the work of a jump protocol up to order $\Gamma_n \tilde{\tau} \rightarrow 0$, regardless of the functional form of the protocol. The result is consistent with the adiabatic work previously found in other systems \cite{Crooks2007}.

For general protocols (finite $\tau \Gamma_n$), the SDE (\ref{LangevinforE2}) must be solved exactly through its propagator, as done in the next section.

\section{The energy propagator}
In this section, we consider the SDE (\ref{LangevinforE2}) and solve its underlying Fokker-Planck equation in order to obtain the propagator $P_t(E|E_0)$. Solvable Fokker-Planck equations for the random energy have been considered before in other contexts \cite{NatPhys2011}. As a matter of fact, a suitable transformation of variables $y_t=E_te^{b(t)}$ with $b(t)=\Gamma_nt-(1/2)\log(\lambda_t/\lambda_0)$ turns (\ref{LangevinforE2}) in a much simpler form:
\begin{equation}
\label{Langevinfory}
dy=(f_n/2)\Gamma_nk_BTdm+\sqrt{2\Gamma_nk_BTy}dW_{m},
\end{equation}
in terms of a new variable $m_t=\int_0^{t}exp(\Gamma_nu)\sqrt{\lambda_0/\lambda_u}du$. Equation above describes a random walk with constant drift and a noise of the type $\sigma(y)\propto\sqrt{y}$. The solution of the underlying Fokker-Planck equation of (\ref{Langevinfory}) is immediate \cite{Paper01}, and using the transformation $P_t(E|E_0)dE=P_{h_t}(y|y_0)dy$, we obtain the nonequilibrium conditional PDF for the energy propagator:

\begin{eqnarray}
\label{PtEE0}
P_t(E|E_0)&=&\alpha_t C_t e^{-C_t(\alpha_tE+E_0 e^{-\Gamma_n t})}\Bigg (\frac{\alpha_tE}{E_0 e^{-\Gamma_0 t}} \Bigg)^{q/2} \times \nonumber \\
&\times& I_q(2C_t\sqrt{\alpha_tE E_0 e^{-\Gamma_n t}}),
\end{eqnarray}
for $E\geq0$ and $E_0\geq0$, where $q=f_n/2-1$, $\alpha_t=(\lambda_0/\lambda_t)^{1/2}$, $C_{t}=\exp(\Gamma_nt)/(\Gamma_nk_BT m_t)$, and $I_q$ is the modified Bessel function of the first kind \cite{Arfken2012}. 
As a straightforward application of (\ref{PtEE0}), we may suppose the particle is initially found in equilibrium with a reservoir of temperature $T_0$. In this case, the initial PDF for the energy is Maxwell-Boltzmann (MB), $P_0(E)=\beta_0^{1+q}/\Gamma(q+1)E^{q}e^{-\beta_0 E}$,
with $\beta_0=(k_BT_0)^{-1}$. For $t>0$, the system undergoes a protocol $\lambda_t$ in thermal contact with a reservoir at temperature $T$ for a finite time interval $[0,t]$. In this case, one may write the nonequilibrium energy distribution as the superposition of Eq.~(\ref{PtEE0}) over the initial conditions (MB), using
\begin{equation}
P_t(E)=\int_{0}^{\infty}P_t(E|E_0)P_{0}(E_0)dE_0,
\end{equation}
which results in another MB distribution, $P_t(E)=\beta_t^{1+q}/\Gamma(q+1)E^{q}e^{-\beta_t E}$, with $\beta_t=(k_B T_t)^{-1}$, for a time dependent effective temperature $T_t$. As the nonequilibrium distribution is MB, one obtains $T_t$ simply taking the ensemble average of the SDE (\ref{LangevinforE2}), using $\langle E \rangle = (f_n/2) k_B T_t$, which results in a ODE for $T_t$ given by
\begin{equation}
\frac{dT_t}{dt}=-\Gamma_n ((1-\frac{\dot{\lambda_t}}{2\lambda_t\Gamma_n})T_t-\frac{f_n}{2}k_B T),
\end{equation}
which can be solved easily, for initial condition $T_0$ and $\lambda_0$ at $t=0$, resulting in
\begin{equation}
\label{Teff}
\frac{T_t}{T}=e^{-\Gamma_n t}\sqrt{\frac{\lambda_t}{\lambda_0}}\Bigg(\frac{T_0}{T}+\Gamma_n\int_0^{t}e^{\Gamma_n u}\sqrt{\frac{\lambda_0}{\lambda_u}}du\Bigg).
\end{equation}
Depending on the protocol $\lambda_t$, the effective temperature takes different forms. For example, it is immediate to check that the constant protocol, $\lambda_t=\lambda_0$ for all $t$, leads to the effective temperature $T_t=T+(T_0-T)e^{-\Gamma_n t}$, which represents the thermal relaxation of a system initially prepared at temperature $T_0$ and placed in thermal contact with a reservoir at temperature $T$. Another example is the adiabatic protocol, with $\lambda_t$ going from $\lambda_0$ to $\lambda_\tau$ in a very brief time interval, $\tau\Gamma_n\ll1$. In this adiabatic case, the effective temperature (\ref{Teff}) results in the relation $T_{\tau}/T=\sqrt{\lambda_\tau/\lambda_0}$ which can be directly related to the polytropic equation, $T_0V_0^{2/f_n}=T_\tau V_\tau^{2/f_n}$, when associating an effective volume $V\propto \lambda^{-f_n/4}$ (see Section VI), already obtained from previous nonequilibrium approaches \cite{Crooks2007} for the case $n=1$. General isothermal protocols result in nontrivial time dependent effective temperatures (\ref{Teff}). 
The average work in such cases can be deduced as follows. First, notice that it follows from the nonequilibrium MB distribution, $P_t(E)$, that $\langle E_t\rangle=(f_n/2)k_BT_t$. The expression for $\langle E_t \rangle$ is useful for calculating the ensemble average of the work increment, $\langle dW \rangle$, from (\ref{diffW}).
Therefore, we find the average work, $\langle W_\tau \rangle$, over the interval $[0,\tau]$, taking the ensemble average of (\ref{diffW}) and integrating in time,
\begin{equation}
\label{Workavg}
\langle W_\tau \rangle = \int_0^{\tau}\langle dW \rangle=\frac{f_n}{2}\frac{k_BT}{2}\int_0^{\tau}\frac{\dot{\lambda_t}}{\lambda_t}\frac{T_t}{T} dt.
\end{equation}
For completeness, the average heat is expressed in a similar form:
\begin{equation}
\label{Heatavg}
\langle Q_\tau \rangle = -\frac{f_n}{2}\Gamma_n k_BT\int_0^{\tau}\Bigg(\frac{T_t}{T}-1\Bigg)dt.
\end{equation}
Equations (\ref{Teff}), (\ref{Workavg}) and (\ref{Heatavg}) represent the main algorithm for calculating the average values of work and heat in any given isothermal process. In summary, for a given protocol $\lambda_t$, one finds the effective temperature using (\ref{Teff}) and use it to compute the averages work (\ref{Workavg}) and heat (\ref{Heatavg}). 

In the next section, we show that the model described by Eq. (\ref{LangevinforE2}) satisfies the Crooks fluctuation theorem.

\section{Crooks Fluctuation Theorem}
In the absence of protocol, the SDE (\ref{LangevinforE2}) describes solely the heat dynamics, which satisfies a heat exchange fluctuation theorem \cite{Paper01}, when two reservoirs are considered. In the presence of a protocol and a single reservoir, we show that (\ref{LangevinforE2}) satisfies the Crooks Fluctuation Theorem \cite{Crooks1999} (CFT), which is stated as follows:
\begin{equation}\label{CrooksA}
    \frac{P_t^{A\rightarrow B}(W)}{P_t^{B\rightarrow A}(W)}=e^{\beta (W-\Delta F)},
\end{equation}
where $W$ is the work and $\Delta F$ is the variation of free energy from $A$ to $B$ (points in the phase space). One of the consequences of (\ref{CrooksA}) is the Jarzinski Equality (JE):
\begin{equation}\label{JE}
    \langle e^{-\beta W} \rangle = e^{-\beta \Delta F},
\end{equation}
where the average is taken over all possible trajectories in the phase space conecting points $A$ and $B$. 

In order to show that our model obeys CFT, notice that the approximate work is given its definition from the main text, which is the nonequilibrium (time dependent) stochastic work. In a discrete version, assuming $N$ time steps of size $\tau$, one could write:
\begin{equation}\label{WorkDiscrete}
    W_t=\sum_{i=1}^{N}\frac{\dot{\lambda_i}}{\lambda_i}\frac{E_i}{2}\tau,
\end{equation}
for $t=N\tau$. As the work is given by a sum of stochastic energies, its value depends on the realization of a trajectory of the energy (single dimension), which obeys the SDE (\ref{LangevinforE2}).
Therefore, we start the demonstration considering a single trajectory, $\gamma=(E({t_0}),...,E(t_{N}))=(E_0,...,E_N)$, with a given protocol $\lambda=(\lambda_0,...,\lambda_N)$. The probability of such trajectory is given by the Bayes theorem:
\begin{equation}\label{Probtrajectory}
    P(\gamma)=P(E_0)\prod_{i=1}^{N}P_{\tau}(E_{i+1}|E_i)dE_{i+1},
\end{equation}
where $P_\tau(E_{i+1}|E_i)$ is computed bellow considering the discrete time $t=n\tau$. Notice that, the ratio between $P(\gamma)$ and the probability of the backward trajectory, $P(\gamma')$, with $\gamma'=(E_N,...,E_0)$ and $\lambda'=(\lambda_N,...,\lambda_0)$, can be calculated explicitly:
\begin{equation}\label{ratiogamma}
    \frac{P(\gamma)}{P(\gamma')}=\frac{P(E_0)}{P(E_n)}\prod_{i=1}^{N}\frac{P_{\tau}(E_{i+1}|E_i)}{P_{\tau}(E_{i}|E_{i+1})}.
\end{equation}
Therefore, the trajectory depends on the probabilities of small steps $E_i\rightarrow E_{i+1}$. In this case, the SDE (\ref{LangevinforE2}) for the forward process reads:
\begin{eqnarray}\label{SDEdiscreteF}
\Delta E_i=(-\Gamma_n+\frac{\Delta \lambda_i}{2\lambda_i\tau})E_i\tau+\frac{f_n}{2}\Gamma_nk_BT\tau+\\
\sqrt{2\Gamma_0k_BTE_i}\Delta B_\tau,
\end{eqnarray}
where $\Delta E_i=E_{i+1}-E_i$ and $\Delta \lambda_i=\lambda_{i+1}-\lambda_i$. The increment $\Delta B_i$ is gaussian with zero mean and variance $\tau$. For the backwards process, one obtains analogously
\begin{eqnarray}\label{SDEdiscreteB}
\Delta E_{i+1}=(-\Gamma_n-\frac{\Delta \lambda_i}{2\lambda_i\tau})E_{i+1}\tau+\frac{f_n}{2}\Gamma_n k_BT\tau+\\
\sqrt{2\Gamma_nk_BTE_{i+1}}\Delta B_\tau,
\end{eqnarray}
Where we kept terms in order $\tau$ in the drift term $(\Delta \lambda_{i}/\lambda_{i+1}\approx\Delta \lambda_i/\lambda_{i}+\mathcal{O}\tau)$. For clarity, define the drift terms $\mu$ and $\mu'$ above such that $\Delta E_i=\mu+\sigma\Delta B_i$ and $\Delta E_{i+1}=\mu'+\sigma\Delta B_i$, with $\sigma^2=2\Gamma_n k_BT$. For small $\tau$, the increment of both forward (\ref{SDEdiscreteF}) and backwards (\ref{SDEdiscreteB}) are gaussian and we obtain the transition probabilities:
\begin{eqnarray}\label{PtEE0discreteF}
P_\tau(E_{i+1}|E_i)&=&\frac{1}{\sqrt{2\pi E_i \sigma^2\tau}}\exp{\frac{-(\Delta E_i-\mu)^2}{2E_i\sigma^2\tau}},
\end{eqnarray}
\begin{eqnarray}\label{PtEE0discreteB}
P_\tau(E_{i}|E_{i+1})&=&\frac{1}{\sqrt{2\pi E_{i+1}\sigma^2\tau}}\exp{\frac{-(\Delta E_i+\mu')^2}{2E_{i+1}\sigma^2\tau}},
\end{eqnarray}
for the forward and backward processes respectively. Upon replacing the distributions for the forward and backward processes (\ref{PtEE0discreteF}) and (\ref{PtEE0discreteB}) in (\ref{ratiogamma}), one obtains:
\begin{equation}\label{ratiogamma2}
    \frac{P(\gamma)}{P(\gamma')}=\sqrt{\frac{E_{0}}{E_N}}\frac{P(E_0)}{P(E_N)}\exp(\sum_{i=1}^{N}D_i),
\end{equation}
with $D_i$ found as:
\begin{equation}\label{Di}
    D_i=-\beta\Delta E_i +\frac{f_n}{2}\frac{\Delta E_i}{E_i}-\frac{f_n}{4}\frac{\Delta \lambda_i}{\lambda_i}+\beta E_i\frac{\Delta \lambda_i}{2\lambda_i}+\mathcal{O}\tau^2.
\end{equation}
Performing the sum from $i=0$ to $N$, notice that the first term will cancel out in the sum, resulting in the Boltzmann factors $\beta E_n$ and $\beta E_0$. The second and third terms are of the type $\Delta \log E_i$ and $\Delta \log \lambda_i$, summing up to $(f_n/2)\log(E_N/E_0)$ and $(f_n/4)\log(\lambda_N/\lambda_0$), respectively. And summing the last term results in the definition of work (\ref{WorkDiscrete}). Therefore, the ratio (\ref{ratiogamma2}) yields:
\begin{equation}\label{ratiogamma3}
   \frac{P(\gamma)}{P(\gamma')}=\frac{P_0(E_0)}{E_0^q e^{-\beta E_0}}\frac{E_N^q e^{-\beta E_N}}{P_N(E_N)}e^{\beta(W-\Delta F)},
\end{equation}
where $W$ is the average work (\ref{WorkDiscrete}) and $\Delta F$ is the variation of free energy in the forward trajectory ($A \rightarrow B$), $\Delta F = (f_n/4)k_B T\log(\lambda_B/\lambda_A$). Notice that the terms of the type $\propto E^q e^{-\beta E}$ are MB distributions. Therefore, this FT takes into account the ratio between initial distributions (possibly nonequilibrium) and equilibrium ones. Finally, considering the probability of finding the work $W$ in a general process from $A$ to $B$, one needs to sum over all possible trajectories (since they are independent) and use (\ref{ratiogamma3}) to find:
\begin{equation}\label{WorkFT}
    \frac{P_t^{A\rightarrow B}(W)}{P_t^{B\rightarrow A}(W)}=\frac{P_0(E_0)}{P_{MB}(E_0)}\frac{P_{MB}(E_N)}{P_N(E_N)}e^{\beta (W-\Delta F)}.
\end{equation}
At this point, is worth noticing that (\ref{WorkFT}) resembles the fluctuation theorem found in \cite{GONG15} for nonequilibrium initial states $P_0(E_0)$ and $P_n(E_n)$. In the specific case when both PDFs are in equilibrium with the reservoir (MB distributions), the first two fractions cancel out and the CFT (\ref{CrooksA}) is obtained.

\section{Optimal Protocol} 
In this section, we are interested in the protocol that minimizes the average work (\ref{Workavg}), with boundary conditions $\lambda_0$ at $t=0$ and $\lambda_\tau$ at $t=\tau$, for a system initially in equilibrium with the reservoir ($T_0=T$). Moreover, we are interested in a time interval $\tau$ such that $\tau \Gamma_n > 0$ for the protocol, which makes dissipation effects relevant. We point out that a setup with different boundary conditions was recently solved for the harmonic case \cite{Dechant2017}. 
It is possible to go beyond the average work, $\langle W \rangle$, and optimize other statistical moments. For instance, in the overdamped limit, a combined optimization of mean and variance leads to intriguing phase transitions in the protocol space \cite{Alexandre2018}. A general optimization strategy requires knowledge of the moment generating function $\langle e^{sW} \rangle$. For simplicity, we limit the scope of the presentation to the optimization of the average work, as it results in interesting protocols with notable applications in heat engines \cite{Dechant2017}.

\subsection{Average work functional}
We start by considering $\lambda_\tau>\lambda_0$ without loss of generality. This is the nonequilibrium analogue of a isothermal compression. The average work is defined using (\ref{Workavg}), where the time dependent temperature is given by (\ref{Teff}), making $T_0=T$:
\begin{equation}
\label{Temp}
T_t=e^{-\Gamma_n t}\sqrt{\frac{\lambda_t}{\lambda_0}}\Bigg(1+\Gamma_n\int_0^{t}e^{\Gamma_nu}\sqrt{\frac{\lambda_0}{\lambda_u}}du\Bigg).
\end{equation}
Fortunately, the maximization the functional (\ref{Workavg}) with (\ref{Temp}) is feasible analytically. Define auxiliary variables $h_t=1+\Gamma_n\int_0^t e^{\Gamma_n u}\sqrt{\lambda_0/\lambda_u}du$, such that $\dot{h_t}=\Gamma_ne^{\Gamma_n t}\sqrt{\lambda_0/\lambda_t}$ and one can show that
\begin{equation}
\label{ddoth}
\frac{\ddot{h_t}}{\dot{h}_t}=
\Gamma_n-\frac{1}{2}\frac{\dot{\lambda_t}}{\lambda_t}.
\end{equation}
Using (\ref{ddoth}), the work functional (\ref{Workavg}) is rewritten in terms of $h_t$ and its derivatives
\begin{equation}
\label{Workfunctional2}
\langle W_\tau \rangle=
\frac{f_n}{2}k_BT\int_0^{\tau}\big(1-\frac{\ddot{h_t}}{\Gamma_0\dot{h}_t}\big)
\frac{\Gamma_n h_t}{\dot{h_t}}\Gamma_n dt,
\end{equation}
with the auxiliary variable $h_t=1+\Gamma_n\int_0^t e^{\Gamma_n u}\sqrt{\lambda_0/\lambda_u}du$, with boundary conditions $h_0=1$, $\dot{h_0}=\Gamma_0$, and  $\dot{h_\tau}=\Gamma_ne^{\Gamma_n\tau}\sqrt{\lambda_0/\lambda_\tau}$. Using Euler-Lagrange (EL) equations to minimize (\ref{Workfunctional2}), one obtains 
\begin{equation}
\label{ELfinal}
    \ddot{h}h=\dot{h}^2,
\end{equation}
with solution $h_t=Ae^{bt}$, for constants $A,b$. But the solution fails to satisfy all boundary conditions for finite $\tau$. This situation of broken extrema prevents the optimal protocol to be smooth everywhere \cite{Variations}. It suggests the optimal protocol should include discontinuous jumps in the values of $\lambda$ at $t=0$ and $t=\tau$, as previously obtained for harmonic potentials in different optimization setups \cite{Gomez2008,Dechant2015,Dechant2017}. In this case, the strategy is to consider a function with a discontinuity. First, we split the time interval $[0,\tau]$ in two $[0,\tau_1]$ and $[\tau_1,\tau]$. The EL equation must be satisfied in both intervals. However, for the same reason presented before, there is not a solution for the interval $[0,\tau_1]$, unless the discontinuities are at $t=0$ and $t=\tau$. It means the protocol starts with a kick at $t=0$, taking $\lambda_0$ to a constant $\lambda_1$ instantly (adiabatically), with respect to the thermal timescale ($\Gamma_n^{-1}$). Now the effective temperature starts at $T_0$ and reads
\begin{equation}
\frac{T_t}{T}=e^{-\Gamma_n t}\sqrt{\frac{\lambda_t}{\lambda_0}}\Bigg(\sqrt{\frac{\lambda_1}{\lambda_0}}+\Gamma_n\int_0^{t}e^{\Gamma_n u}\sqrt{\frac{\lambda_0}{\lambda_u}}du\Bigg),
\end{equation}
where we replaced $T_0/T=\sqrt{\lambda_1/\lambda_0}$, from (\ref{adiabaticenergy}). The final condition $\lambda_2$ is also a free parameter (letting to the second jump $\lambda_2\rightarrow \lambda_\tau$ at $t=\tau$). Now, solving (\ref{ELfinal}) with new boundary conditions allows one to write the average work in terms of a single parameter $\lambda_1$. The optimal protocol starts with a jump $\lambda_0\rightarrow \lambda_1$, followed by a smooth exponential part,
\begin{equation}
\label{smooth}
\lambda_t=\lambda_1(\lambda_2/\lambda_1)^{t/\tau},
\end{equation}
where this smooth part is in close analogy to the particle in a box \cite{GONG16}. Finally, a second jump takes place, $\lambda_2\rightarrow\lambda_\tau$. The jumps have a defined values obtained from the bondary conditions.

\subsection{Obtaining the jumps}

Both jumps are related from the the boundary conditions of the problem,
\begin{equation}
\label{BC}
1-\frac{1}{\Gamma_n\tau}
\log {\sqrt{\frac{\lambda_2}{\lambda_1}}}
=\sqrt{\frac{\lambda_0}{\lambda_1}},
\end{equation}

\begin{figure}[ht]
\includegraphics[width=3.3 in]{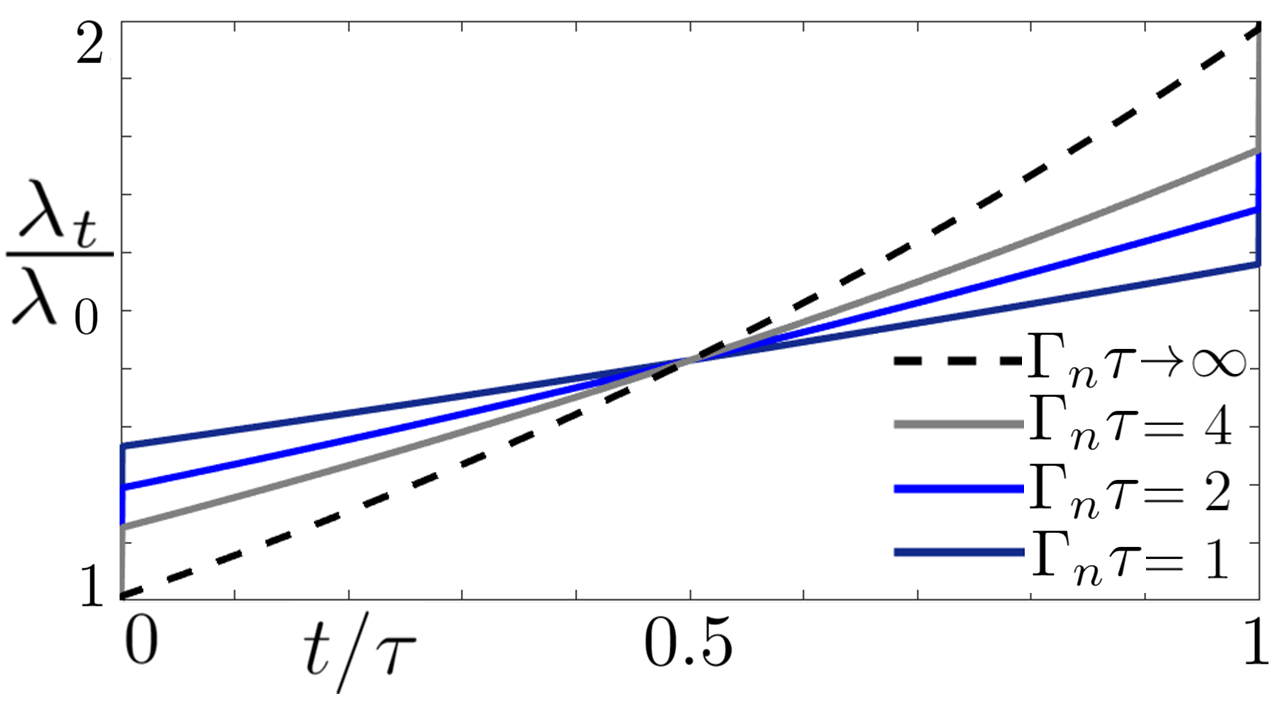}
\caption{(Color online) The optimal protocol, $\lambda_t/\lambda_0$, as a function of time in the isothermal compression for boundary condition $\lambda_\tau/\lambda_0=2$ and different total time duration $\Gamma_n\tau=\{1,2,4,\infty\}$. For all single-well potentials ($n=1,2,...$), the protocols are composed of jumps (at $t=0$ and $t=\tau$) connected by a smooth part, $\lambda_t=\lambda_1(\lambda_2/\lambda_1)^{t/\tau}$. Notice the smaller interval $\Gamma_n\tau=1$ (dark blue) results in a higher jump. Intermediate cases, $\Gamma_n\tau=2$ (blue) and $\Gamma_n\tau=4$ (grey), decrease the
magnitude of the jumps. The long range limit $\Gamma_n\tau\rightarrow\infty$ (dashed) is purely exponential in time.}
\label{fig2}
\end{figure}

and the average work (\ref{Workavg}) can be written in terms of $\lambda_1$ as follows
\begin{equation}
\label{Workfinal}
\frac{\langle W_\tau \rangle}{(f_n/2)k_BT} =
\Big(\Gamma_n\tau\big(\sqrt{\frac{\lambda_1}{\lambda_0}}-1\big)+\sqrt{\frac{\lambda_\tau}{\lambda_0}}e^{\Gamma_n\tau\sqrt{\frac{\lambda_0}{\lambda_1}}-1}-1\Big),
\end{equation}
which now can be minimized with respect to $\lambda_1$, leading to the closed form expression for the jump

\begin{equation}
\label{jump}
\frac{\lambda_1}{\lambda_0}=
\frac{(\Gamma_n \tau/2)^2}{\mathcal{W}((\lambda_0/\lambda_\tau)^{1/4}e^{\Gamma_n \tau /2}\Gamma_n\tau/2))^2},
\end{equation}
where $\mathcal{W}$ is the Lambert $W$ function.
Finally, the protocol takes a second jump $\lambda_2\rightarrow \lambda_\tau$ at $t=\tau$. From (\ref{BC}), the final condition for the smooth part of the protocol, $\lambda_2$, satisfies the relation
\begin{equation}
\label{secondjump}
\lambda_2/\lambda_\tau=\lambda_0/\lambda_1,
\end{equation}
where $\lambda_2$ is found in terms of $\lambda_1$ from (\ref{jump}). We show the optimal protocol $\lambda_t$ as a function of time using (\ref{smooth}) in Fig. 1, with jumps $\lambda_1$ and $\lambda_2$ given by equations (\ref{jump}) and (\ref{secondjump}) respectively, for different values of $\Gamma_n\tau$ and $\lambda_\tau/\lambda_0=2$, valid for all values of $n$.
We also use the magnitude of the first jump $\lambda_1$ from (\ref{jump}) to compute the optimal work using (\ref{Workfinal}), depicted in Fig. 2, for different values of $\Gamma_n \tau$ and $\lambda_\tau/\lambda_0$. We remark that the optimal protocol obtained in \cite{GONG16} does not contain adiabatic jumps. This is consistent with the slow expanding considered in the paper, as the jumps tend to zero in the limit $\tau \rightarrow \infty$.

\subsection{Interpretation of the optimal protocol}

\begin{figure}[ht]
\includegraphics[width=3.3 in]{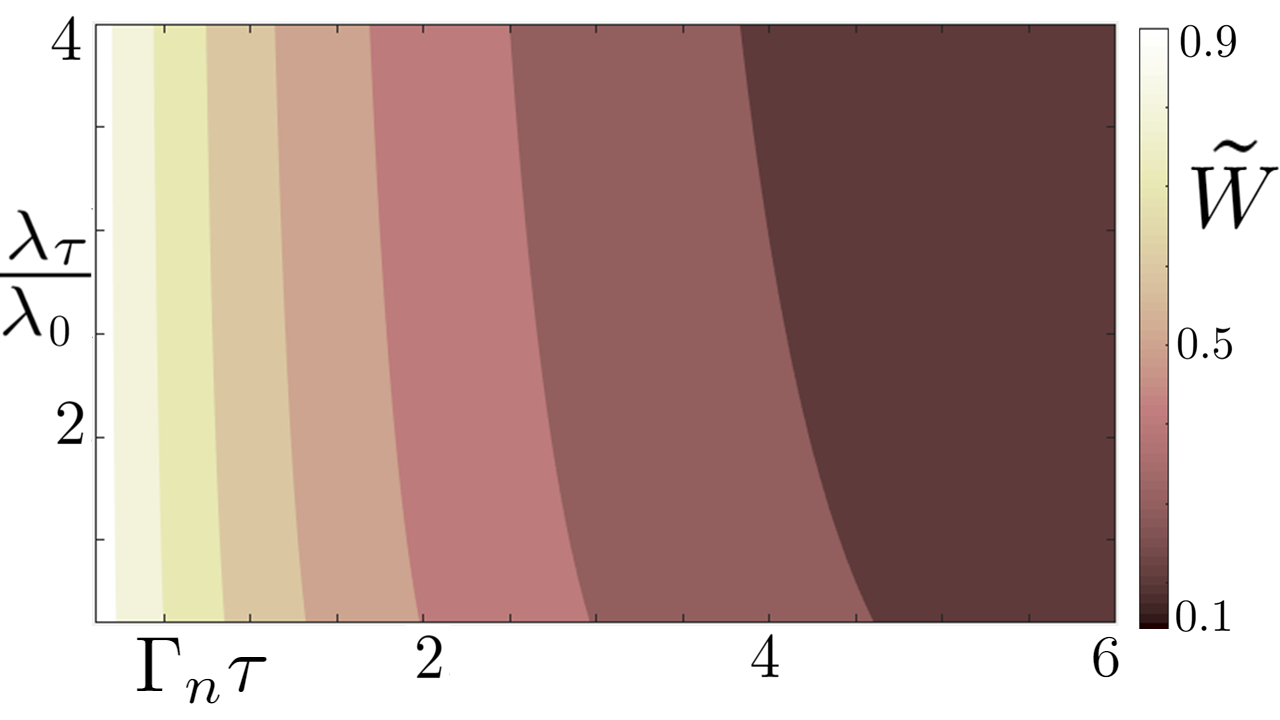}
\caption{(Color online) The normalized optimal work in the isothermal process, $\tilde{W}=(\langle W_\tau \rangle-\Delta F)/(W_a-\Delta F)$,  for all single-well potentials, where $W_\tau$ is the optimal work (\ref{Workfinal}), $W_a$ is the adiabatic (maximum) work and $\Delta F$ is the free energy difference (minimum work). Notice that $\Gamma_n\tau\rightarrow 0$ makes the optimal work close to the adiabatic value ($\tilde{W}\rightarrow 1$), as expected for a fast compression. However, for a long time, $\Gamma_n\tau\rightarrow \infty$, the optimal work approaches the free energy difference ($\tilde{W}\rightarrow 0$), as expected in a quasistatic process.}
\label{fig1}
\end{figure}
In this section, it was found that the compression protocol that minimizes the dissipated work for a fixed time duration $\tau$ and bounded values for the trapping parameters $\lambda_0$, $\lambda_{\tau}$ is given by an initial fast adiabatic compression, followed by an exponential time-dependent isothermal drive, and ending with another fast adiabatic jump. 

Intuitively, one might understand this optimal protocol as follows: the system needs to be artificially ``heated" through a jump $\lambda_0\rightarrow\lambda_1$ to a temperature $T_0>T$ at $t=0$. After that, at a higher effective temperature, the heat dissipation and the energy pumped into the system by the protocol are compensated for the smooth exponential part, keeping the effective temperature constant at $T_1$ for $(0,\tau)$. Similarly to the quasistatic case $\Gamma_n\tau\rightarrow \infty$ where $T_t=T$ during the whole process, the far from equilibrium optimal protocol (finite $\Gamma_n\tau$) still keeps the constant effective temperature condition but now at a temperature $T_t=T_1$ created artificially by the first adiabatic jump. As the system is described by a MB distribution through the whole process, it means the average energy is constant during ($0,\tau$) and the work (\ref{Workavg}) increases linearly with time (see Fig.~3).
Also notice in (\ref{Workfinal}) that the full adiabatic case is recovered, $\langle W_\tau \rangle = (f_n/2)k_BT(\sqrt{\lambda_\tau/\lambda_0}-1)$ in the limit $\Gamma_n\tau\rightarrow 0$, as expected. Alternatively, the long range limit, $\lambda_1\rightarrow\lambda_0$ and $\lambda_2\rightarrow\lambda_\tau$, results in $\langle W_\tau \rangle = \Delta F + \Sigma/\tau$, with the free energy $\Delta F = (f_n/2)k_BT\log\sqrt{\lambda_\tau/\lambda_0}$ and $\Sigma=(f_n/2) k_B T (1/4\Gamma_n)\log(\lambda_\tau/\lambda_0)^2\geq 0$, which is related to the complementarity relation already suggested in overdamped systems \cite{Sekimoto1997}. 
As a comparison, the pure exponential protocol (without jumps, ie, $\lambda_1=\lambda_0$ and $\lambda_2=\lambda_\tau$) can also be calculated. Using $\lambda_t=\lambda_0 (\lambda_\tau/\lambda_0)^{t/\tau}$ in (\ref{Teff}), it results in the effective temperature
\begin{equation}
\label{Teff2}
\frac{T_t}{T}=(1/g_\tau)+(1-(1/g_\tau))e^{-g_\tau\Gamma_n t},
\end{equation}
where $g_\tau\neq0$ is defined as $g_\tau=[1-\frac{1}{2\Gamma_n\tau}\log{(\lambda_\tau/\lambda_0)}]$,

\begin{figure}[ht]
\includegraphics[width=3.3 in]{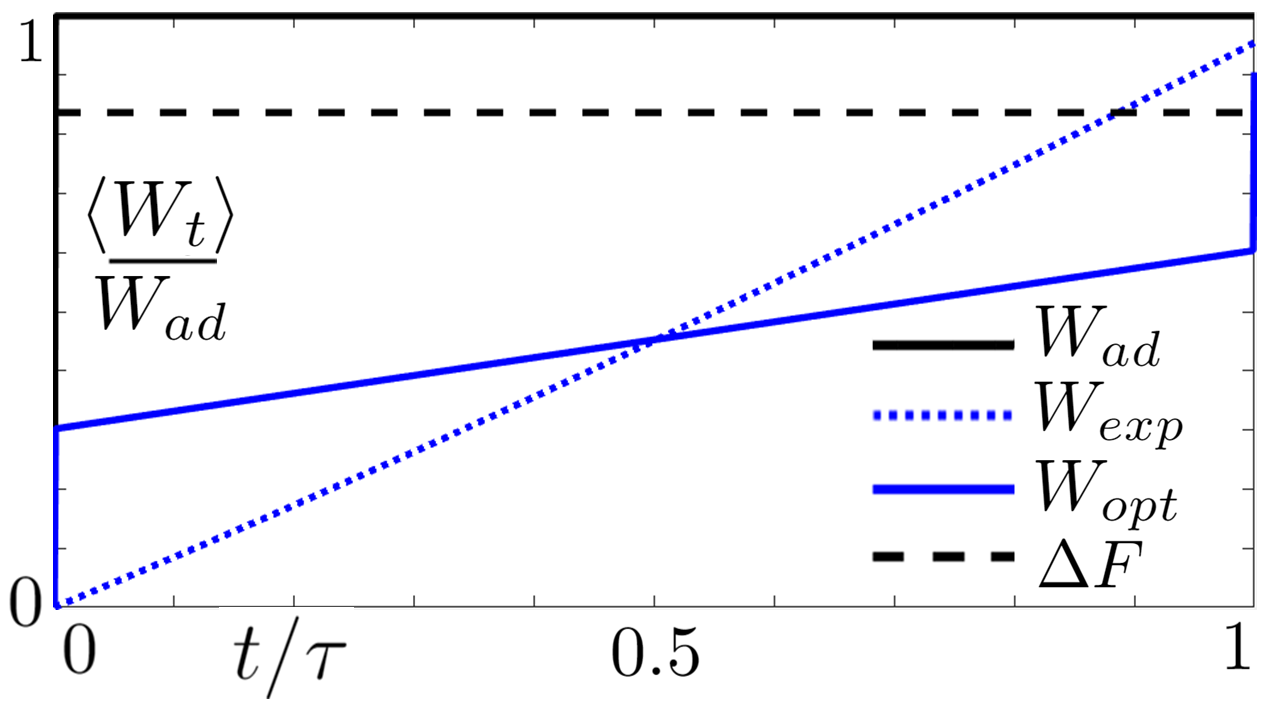}
\caption{(Color online) The average work as a function of time for $\Gamma_n\tau=1$ and $\lambda_\tau/\lambda_0=2$. The optimal protocol is compared (solid blue) to a purely exponential protocol (dotted blue). The adiabatic protocol is depicted in solid black and the free energy is dashed. Notice the optimal protocol starts ahead when compared to the exponential, but the initial jump allows the system to keep a constant effective temperature for $t>0$, which results in a work increasing linearly in time. In the end, the optimal protocol performs the second jump, leaving the final average work below the purely exponential protocol, as expected. The difference between their final values (at $t=\tau$) and the free energy is the irreversible work.}
\label{fig3}
\end{figure}
for $0\leq t \leq\tau$. 
The averages work are computed inserting (\ref{Teff2}) in (\ref{Workavg}),
\begin{equation}
\label{Workexpo}
\frac{\langle W_t \rangle}{(f_n/2)k_BT} = ((1/g_\tau)-1)(\Gamma_n t-((1/g_\tau)-1)(1-e^{-g_\tau\Gamma_n t}).
\end{equation}
For $g_\tau=0$, one obtains a simple form for the effective temperature, $T_t/T=\Gamma_n t +1$, for which the average work follows immediately as in the previous case.
The work of the optimal protocol is compared to the exponential protocol (\ref{Workexpo}) as a function of time in Fig.~3. We remark that the adiabatic protocol (also called jump) seems instantaneous in Fig.~3 for the timescale $\tau$, but it actually takes a finite time $\tilde{\tau}\ll \tau$ defined in Sec. II. D, which makes $\tilde{\tau}/\tau \approx 0$ in Fig. ~3.

\section {Analogy to a gas of free brownian particles}
We now show that it is possible to obtain a model for independent free Brownian particles by using kinetic theory considerations and which is equivalent to (\ref{LangevinforE2}). We start by setting $k_t=0$ for all $t\geq0$ in (\ref{Langevin}), thus obtaining the Brownian motion with energy (per particle) given by $E=\sum_{i=1}^{f}p_{i}^2/(2m)$, where $f$ denotes the number of degrees of freedom. It follows from Ito calculus  \cite{Paper01} that the SDE for the energy increment is given by (\ref{LangevinforE2}) with $dE=dQ$ and $dW=0$, and it accounts solely for heat exchange. In order to perform work over the system, one must change the available volume $V$ of the gas. The system's energy in this case is expected to behave as in the first law $dE=dQ+dW$, with $dW=-PdV$ in our notation, where $P$ is the gas pressure. Now we assume perfect elastic collision with the walls, so the classic kinetic theory can be used to write the stochastic pressure as a function of the kinetic energy $E$ as the usual identity $P=(2/f)E/V$, in $f$ degrees of freedom. This approach yields to the expression for the increment of work of the Brownian gas $dW=-E(2/f)(\dot{V}_t/V_t)dt$. It is worth noting that, using the single-well potential work (\ref{diffW}) and the effective volume as $V \propto \lambda^{-f/4}$ results in $\dot{\lambda_t}/(2{\lambda_t})=-(2/f)\dot{V_t}/V_t$. Finally, upon combining the increments of heat and work, one obtains the same form of the SDE (\ref{LangevinforE2}), proving the analogy between the highly underdamped limit of the Langevin system and a gas of Brownian particles. Therefore, all the results derived for the SDE (\ref{LangevinforE2}) also apply to the Brownian gas: the Newton's law of cooling and the heat fluctuation theorem \cite{Paper01}, the propagator for the stochastic energy $P_t(E|E_0)$ from Eq. (\ref{PtEE0}), Crooks FT (\ref{WorkFT}), the nonequilibrium MB distribution with effective temperature given by Eq. (\ref{Teff}), and the optimal work protocol found in Eq. (\ref{Workavg}). It is interesting to notice that the same exponential optimal work protocol of Eq.(\ref{Workavg}) was rigorously found for the linear regime in the case of a single particle in a box \cite{GONG16}, but without the adiabatic jumps (\ref{jump}) and (\ref{secondjump}). Moreover, the box could also be modeled from (\ref{Langevin}) by making the limit $n\rightarrow\infty$, which in turn makes the potential $U(x)=0$ for $|x|<L$ and $U(x)\rightarrow \infty$ otherwise, leading us to the same conclusions. The difference between the current approach and the particle in a box \cite{GONG16} is the absence of jumps, possibly due to the slow protocol requirement assumed in their treatment.

\section{Conclusions and Perspectives} Usually in thermodynamics, processes are either too fast (e. g. adiabatic) with $\Gamma_0\tau\rightarrow 0$, or too slow (quasistatic) with $\Gamma_0\tau\rightarrow\infty$. A solvable stochastic thermodynamics framework as (\ref{LangevinforE2}) has the advantage of providing a description of different nonequilibrium processes for finite time intervals. In this paper, we consider a classical particle submitted to a generalized single-well potential in high vacuum and derive the time dependent probability density function for the energy (\ref{PtEE0}) explicitly, which allows the computation of far from equilibrium thermodynamics quantities. As a matter of fact, in our system the nonequilibrium information is encoded in a general time dependent effective temperature (\ref{Teff}) and fractional degrees of freedom $f_n$. We showed the system satisfies the Crooks FT on its more general form. In addition, as a relevant application, we have found the optimal protocol $\lambda_t$, for fixed values of $\lambda_0$ and $\lambda_\tau$, that produces minimum average work over a finite time window $[0,\tau]$. The optimal protocol always has adiabatic jumps (at $t=0$ and $t=\tau$) and a smooth exponential part (for $t>0$) for all kinds of single-well potentials. This finding sheds light on the analytic description of general thermal engines, as discontinuous protocols are likely to appear \cite{Dechant2017}. Notice that the adiabatic, isochoric and isothermal processes are important parts of the  description of nanoscopic thermal engines. The calculations of power and efficiency for different sorts of protocols requires dealing with the nonequilibrium thermodynamics observable quantities obtained in this paper. Other types of work optimization, such as a combination of average and variance of work \cite{Alexandre2018}, might be carried in terms of the moment generating function $\langle e^{sW} \rangle$, where different optimal protocols are expected. This type of optimization  goes beyond the scope of this paper and it is left for future research. As a final remark, it is relevant to mention that the highly underdamped limit SDE for the energy (\ref{LangevinforE2}) presents a form that resembles the thermodynamics of free Brownian particles, with similar exponential optimal protocols for long time isothermal processes \cite{GONG16}. 


\begin{thebibliography}{99}


\bibitem{Seifert2012}U. Seifert, Rep. Prog. Phys. {\bf 75}, 126001 (2012). 

\bibitem{Bustamante2005}C. Bustamante, J. Liphardt, and F. Ritort, Phys. Today {\bf 58}, 43 (2005). 


\bibitem{Sekimoto2010} K. Sekimoto,  {\it Stochastic Energetics} (Springer, Berlin, 2010).

\bibitem{Aspelmeyer2014} 
M. Aspelmeyer, T. J. Kippenberg, and F. Marquardt,
Rev. Mod. Phys. \textbf{86}, 1391 (2014).

\bibitem{Gieseler2018} J. Gieseler and J. Millen, Entropy, \textbf{20} 326 (2018).

\bibitem{Liphardt2002} J. Liphardt, S. Dumont, S. B. Smith, I. Tinoco, and C. Bustamante, Science \textbf{296}, 1832 (2002).
\bibitem{Collin2005} D. Collin, F. Ritort, C. Jarzynski, S. B. Smith, I. Tinoco, and C. Bustamante, Nature 437, 231 (2005).
\bibitem{Alemany2012} A. Alemany, A. Mossa, I. Junier, and F. Ritort, Nature Phys. 8, 688 (2012).
\bibitem{Wang2002} G. M. Wang, E. M. Sevick, E. Mittag, D. J. Searles, and
D. J. Evans, Phys. Rev. Lett. 89, 050601 (2002).
\bibitem{Gavrilov2014}
Y. Jun, M. Gavrilov, and J. Bechhoefer, Phys. Rev. Lett. 113, 190601 (2014).

\bibitem{Jarzinski1997} C. Jarzynski, Phys. Rev. Lett. {\bf 78}, 2690 (1997). 

\bibitem{Yin2013} 
Z. Yin, A. Geraci, T. Li,
Int. J. Mod. Phys. B \textbf{27}, 1330018–1330027 (2013).
\bibitem{Vamivakas2016}
N. Vamivakas, M. Bhattacharya and P. Barker,
Opt. Photon. News \textbf{27} 42–49 (2016).

\bibitem{Gieseler2012}J. Gieseler, R. Quidant, C. Dellago, and L. Novotny, Nature Nanotech. {\bf 9}, 358 (2014). 

\bibitem{Hoang2018}
T. Hoang et al . Phys. Rev. Lett. , \textbf{120} 080602 (2018).

\bibitem{GieselerPRL2012}
J. Gieseler, B. Deutsch, R. Quidant, L. Novotny, L. Phys. Rev. Lett. \textbf{109}  103603 (2012).

\bibitem{Millen2015}
J. Millen, P. Fonseca, T. Mavrogordatos, T. Monteiro, P. Barker, Phys. Rev. Lett. \textbf{114}, 123602 (2015).

\bibitem{Tongcang2011}
T. Li, S. Kheifets and M. Raizen, Nat. Phys. \textbf{7} 527 (2011).

\bibitem{Jain2016} Jain, V.; Gieseler, J.; Moritz, C.; Dellago, C.; Quidant, R.; Novotny, L.  Phys. Rev. Lett. 2016, \textbf{116}, 243601

\bibitem{Geraci2010}
A. Geraci, S. Papp and J. Kitching, Phys. Rev. Lett. \textbf{105}, 101101 (2010).

\bibitem{Bose2017}
S. Bose, A. Mazumdar, G. Morley, H. Ulbricht, M. Toros, M. Paternostro, A. Geraci, P. Barker, S. Kim and G. Milburn, 
Phys. Rev. Lett. \textbf{119} 240401 (2017).

\bibitem{Dechant2015}
A. Dechant, N. Kiesel, E. Lutz, Phys. Rev. Lett. \textbf{114}, 183602 (2015).

\bibitem{Blickle2011}V. Blickle, and C. Bechinger,  Nature Phys. {\bf 8}, 143 (2011). 

\bibitem{Abah2012}
O. Abah, J. Ro\ss{}nagel, G. Jacob, S. Deffner, F. Schmidt-Kaler, K. Singer, and E. Lutz
Phys. Rev. Lett. \textbf{109}, 203006 (2012).

\bibitem{Rossnagel2014} J. Ro\ss{}nagel, O. Abah, F. Schmidt-Kaler., K. Singer, and
E. Lutz, Phys. Rev. Lett. \textbf{112}, 030602 (2014).

\bibitem{Dechant2017} A. Dechant, N. Kiesel and E. Lutz, Europhysics Letters \textbf{119}, 5 (2017).

\bibitem{Alexandre2018}
Alexandre P. Solon and Jordan M. Horowitz Phys. Rev. Lett. 120, 180605

\bibitem{Agarwal2013} G. S. Agarwal and S. Chaturvedi, Physical Review E \textbf{88}, 012130 (2013).

\bibitem{Schmiedl2007}
Tim Schmiedl and Udo Seifert
Phys. Rev. Lett.\textbf{98}, 108301 (2007).

\bibitem{Schmiedl2008} 
T. Schmiedl and U. Seifert, EPL \textbf{81}, 20003 (2008).

\bibitem{Aurell2011}
E. Aurell, C. Mejia Monasterio, P. Muratore-Ginanneschi Physical Review letters 106 (25), 250601 (2011)

\bibitem{Aurell2012}
E. Aurell, L. Gawedzki, C. Mejia Monasterio, R. Mohayae, P. Muratore-Ginanneschi, J. Stat. Phys. \textbf{147}, 487-505 (2012).


\bibitem{Gomez2008}
A. Gomez-Marin, T. Schmiedl, and U. Seifert, J. Chem.
Phys. \textbf{129}, 024114 (2008).

\bibitem{GONG16}
Z. Gong, Y. Lan, and H. Quan
Phys. Rev. Lett. \textbf{117}, 180603 (2016).

\bibitem{Mallick2003}
K. Mallick, P. Marcq, European Physical Journal B, \textbf{31}, 4 (2003).

\bibitem{Mallick2005}
K. Mallick, P. Marcq, Journal of Stat. Phys. \textbf{119}, 1–2 (2005).

\bibitem{Mandrysz2018}
M. Mandrysz, B. Dybiec, arXiv:1810.09186 (2018).



\bibitem{Sekimoto1997} K. Sekimoto, S.I. Sasa, J. Phys. Soc. Jpn. {\bf 66}, 3326 (1997).

\bibitem{Oeksendal2003} B. Oksendal, {\it Stochastic Differential Equations: An Introduction with Applications} (Springer, Berlin, 2003).


\bibitem{NatPhys2011} Bunin, G., D’Alessio, L., Kafri, Y. and
Polkovnikov, A., 2011. Nat. Phys. Physics 7, 913 (2011).

\bibitem{Paper01} D. S. P. Salazar, S. A. Lira, Journal of Phys. A {\bf 49}, 465001 (2016)

\bibitem{Pal2014}
A. Pal and S. Sabhapandit
Phys. Rev. E {\bf 90}, 052116.

\bibitem{Gieseler2019}
Millen J., Gieseler J. (2018) Single Particle Thermodynamics with Levitated Nanoparticles. In: Binder F., Correa L., Gogolin C., Anders J., Adesso G. (eds) Thermodynamics in the Quantum Regime. Fundamental Theories of Physics, vol 195. Springer, Cham



\bibitem{Crooks1999}G. E. Crooks, Phys. Rev. E {\bf 60}, 2721 (1999). 

\bibitem{GONG15} Z. Gong and H. T. Quan, Phys. Rev. E {\bf 92}, 012131 (2015). 

\bibitem{Arfken2012} G. B. Arfken, H. Weber, F. E. Harris, {\it Mathematical Methods for Physicists: A Comprehensive Guide} (Academic Press, 2012).

\bibitem{Crooks2007}G. E. Crooks and C. Jarzynski , Phys. Rev. E {\bf 75}, 021116 (2007). 

\bibitem{Variations} I. M. Gelfand, S. V. Fomin, \textit{Calculus of Variations} (Dover, 2000).





























\end{thebibliography}
\end{document}